\documentclass{elsart5p}

\usepackage{graphicx}
\usepackage{lineno}

\begin{document}
\begin{frontmatter}
\title{Acoustic noise in deep ice and environmental conditions \\
  at the South Pole}

\author{Timo Karg}
\ead{karg@physik.uni-wuppertal.de} for the
\author{IceCube collaboration}
\ead[url]{http://icecube.wisc.edu/}
\address{Bergische Universit\"at Wuppertal, Fachbereich C ---
  Mathematik und Naturwissenschaften, 42097 Wuppertal, Germany}

\begin{abstract}
  To study the acoustic properties of the Antarctic ice the South Pole
  Acoustic Test Setup (SPATS) was installed in the upper part of drill
  holes for the IceCube neutrino observatory.  An important parameter
  for the design of a future acoustic neutrino telescope is the
  acoustic background noise in the ice and its spatial and temporal
  variations. We study the absolute noise level depth profile from
  SPATS data and discuss systematic uncertainties. The measured noise
  is very stable over one year of data taking, and we estimate the
  absolute noise level to be $< 10$~mPa in the frequency range from
  10~kHz to 50~kHz at depths below 200~m. This noise level is of the
  same order of magnitude as observed by ocean based acoustic neutrino
  detection projects in good weather conditions.
\end{abstract}

\end{frontmatter}

\section{Introduction}
\label{sec:introduction}

Hadronic processes at highest energies which are anticipated to take
place in cosmic sources like Active Galactic Nuclei or Gamma-ray
bursts can be studied by observation of the ultra-high energy (UHE)
neutrinos which are produced in these interactions (see
\cite{Becker:2008} for a recent review). Further on, there should be a
guaranteed flux of UHE neutrinos from the interaction of high energy
cosmic ray protons with the cosmic microwave background radiation, the
GZK neutrinos \cite{Beresinsky:1969}. Since the theoretically
predicted neutrino fluxes are very low, large detector masses are
needed to detect UHE neutrinos with significant statistics. The
required sensor density of neutrino telescopes in natural media is
determined by the signal attenuation length. While the attenuation
length for optical \v{C}erenkov light in polar ice is of the order of
100~m, much larger attenuation lengths, $> 1$~km, are predicted for
radio and acoustic waves from available data \cite{Barwick:2005,
  Price:2006}. Both types of signals are emitted from the
electromagnetic and hadronic cascades generated in an UHE neutrino
interaction \cite{Askaryan:1962, Askaryan:1957}. The Antarctic ice
sheet offers the unique possibility to combine these three detection
techniques---optical, radio, and acoustic---into a hybrid
detector. This will largely increase event tagging and background
rejection capabilities for the rare UHE neutrino events.

At the site of the IceCube neutrino observatory two test systems have
been deployed to study the properties of the South Pole ice: the
Askaryan Underice Radio Array (AURA) \cite{Landsman:2007} to measure
the radio properties, and the South Pole Acoustic Test System (SPATS)
\cite{Boeser:2007} to study the acoustic properties. The quantities of
interest are the attenuation length, the sound speed profile which
determines the refraction of the acoustic signal, the origin and
temporal and spatial distribution of transient backgrounds, and the
background noise level which is discussed in this work.

SPATS consists of four instrumented strings that are installed in the
upper part of IceCube holes. Each string comprises seven acoustic
sensors and seven transmitters. Three of the strings (Strings A, B,
and C) were deployed in the austral summer 2006/07 and cover the depth
range from 80~m to 400~m. The array was extended in December 2007 by a
fourth string (String D) with second generation sensors and
transmitters between 190~m and 500~m depth. Horizontal baselines from
125~m to 543~m are covered by the setup.

The absolute level and spectral shape of the continuous noise
determine the threshold at which neutrino induced signals can be
extracted from the background and thus set the lower energy threshold
for a given detector configuration. The continuous noise is monitored
in SPATS through an untriggered read out of all sensor channels for
0.5~s at 200~kHz sampling rate once every hour.

Transient acoustic noise sources which can be misidentified as
possible neutrino candidates in acoustic neutrino telescopes are
studied in a triggered run mode. A threshold trigger in any channel
can be used to read out up to three channels on the same string. This
allows the characterization of transient signals as function of
threshold and frequency content (waveform shape).

\section{Properties of the measured noise}
\label{sec:properties}

SPATS is operating successfully and has been taking data continuously
since its installation and commissioning in January 2007. The
untriggered noise recordings show a Gaussian distribution of the ADC
counts in each of the 73 channels continuously operating during the
considered time interval.

\begin{figure}[ht]
  \centering
  \includegraphics[width=0.5\textwidth]{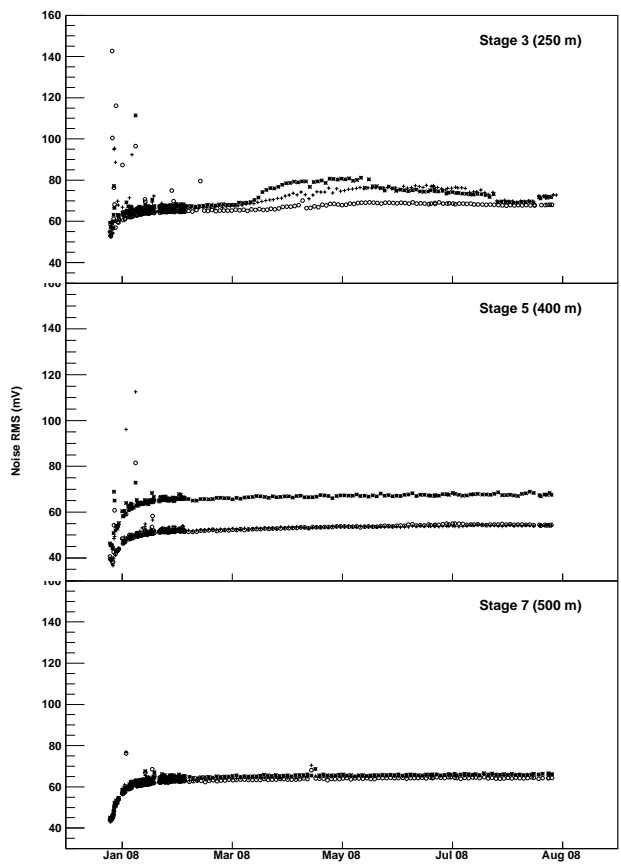}
  \caption{Evolution of the noise RMS for three exemplary sensors from
    String D starting from deployment in December 2007. The three
    different symbols in each panel represent the three sensor
    channels available at each depth. The excess points in the noise
    level in January and February are correlated in time with IceCube
    drilling activity.}
  \label{fig:noise_vs_time}
\end{figure}

Figure \ref{fig:noise_vs_time} shows the RMS of the Gaussian
distribution as a function of time for the whole data taking period of
String D, starting 26 December 2007. It can be seen, that the noise
level stabilizes two to three weeks after deployment in all channels,
with the exception of two channels on Stage 3 which is believed to be
an instrumental problem. The continuous increase of the measured noise
level at the beginning is assumed to be due to an increasing
sensitivity at low temperatures and an improved coupling to the bulk
ice after freeze-in. The typical deviation from the mean noise level
is $\frac{\sigma_{\mathrm{RMS}}}{\langle \mathrm{RMS} \rangle} <
10^{-2}$.

The excess points in the noise data before 25 January 2008 are
correlated in time to the drilling of IceCube holes. The noise level
in each channel rises for few minutes when the drill head passes the
sensor depth level on its way down and up again. The drill noise was
observed up to the furthest hole drilled in the 2007/08 season which
had a horizontal distance of 660~m.

\section{Environmental conditions and absolute calibration}
\label{sec:calibration}

The determination of the absolute noise level from the measured
voltage requires calibration of the sensors. The SPATS sensors have
been calibrated in liquid water at 0$\,^\circ$C in the frequency range
from 10~kHz to 80~kHz \cite{Fischer:2006} prior to
deployment. However, it is unknown how the sensitivity changes when
the sensor is deployed in Antarctic ice, i.e.~under the combined
influence of low temperatures ($-50\,^\circ$C) and high static
pressure. External pressure is produced by the water column inside the
drill hole during deployment and increases during freeze in since the
hole freezes from the top, creating a water filled cavity. After the
hole is frozen completely relaxation of the freshly frozen ``hole
ice'' to the surrounding bulk ice occurs.

\begin{figure}[ht]
  \centering
  \includegraphics[width=0.5\textwidth]{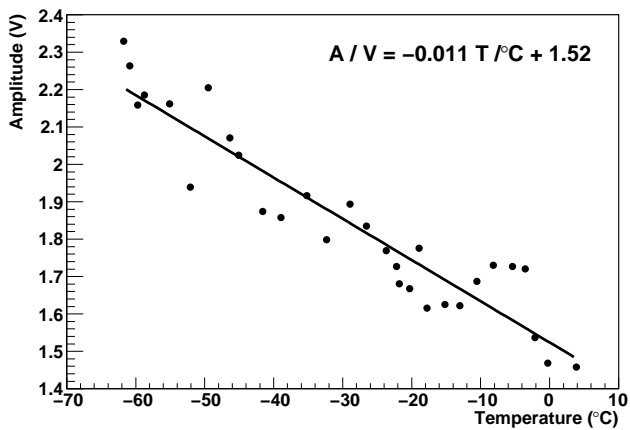}
  \caption{Peak to peak amplitude of a transmitter signal measured in
    air with a SPATS sensor as a function of ambient temperature
    \cite{Bothe:2008} (dots), and linear fit to the data. The measured
    amplitude is used as an estimator for the sensitivity of the
    sensor.}
  \label{fig:sensitivity_vs_temp}
\end{figure}

Lab measurements have shown that the sensitivity of the SPATS sensor
in air at atmospheric pressure increases by a factor of 1.4 when
cooled down from 0$\,^\circ$C to $-50\,^\circ$C (cf.~Figure
\ref{fig:sensitivity_vs_temp}). In this measurement the recorded peak
to peak amplitude of a pulse emitted from a piezoelectric transmitter
driven by a fixed input voltage and at the same ambient temperature as
the receiver is used as an estimator for the sensitivity of the
sensor. We use the value of 1.4 to estimate the unknown change of the
sensor sensitivity deep in Antarctic ice. A measurement of the
sensitivity at room temperature as a function of static pressure was
performed in a pressure vessel. Preliminary results indicate a change
in sensitivity of $< 20$~\% in the pressure range from 1~bar to
100~bar.

An additional source of systematic uncertainty under study is the
influence of the ``hole ice'': the ice that is freshly refrozen in the
IceCube holes after drilling. This ice has a lattice structure
different from the bulk ice and so the propagation and attenuation of
acoustic signals in this ice column is difficult to describe and can
influence the observed noise level. It is expected that under the
ambient pressure a reordering takes place in the ice and the ``hole
ice'' becomes compact bulk ice. The time scale for this process
however is not known yet.

\section{Estimation of the absolute noise level}
\label{sec:absolute_noise}

The properties of the continuous noise can be described best in terms
of the power spectral density (PSD). We chose the following definition
of the PSD of a digitized signal of $N$ samples $X_k (k = 0 \dots N -
1)$ recorded with sampling frequency $f_s$:

\begin{displaymath}
  \mathrm{PSD}_j = \frac{2 \vert \tilde{X}_j \vert^2}{f_s (N - 1)},
  \,\, \textrm{with } \tilde{X}_j = \sum_{k = 0}^{N - 1} X_k e^{-2 \pi
    i \frac{jk}{N}}.
\end{displaymath}

\noindent The coefficient $\mathrm{PSD}_j$ corresponds to a frequency
of $f_j = \frac{f_s}{N} \cdot j, \,\, (j = 0 \ldots N / 2)$. With this
definition the integral over the PSD is equal to the RMS of the signal
in the time domain, bandwidth limited to the chosen frequency range.

\begin{figure}[ht]
  \centering
  \includegraphics[width=0.5\textwidth]{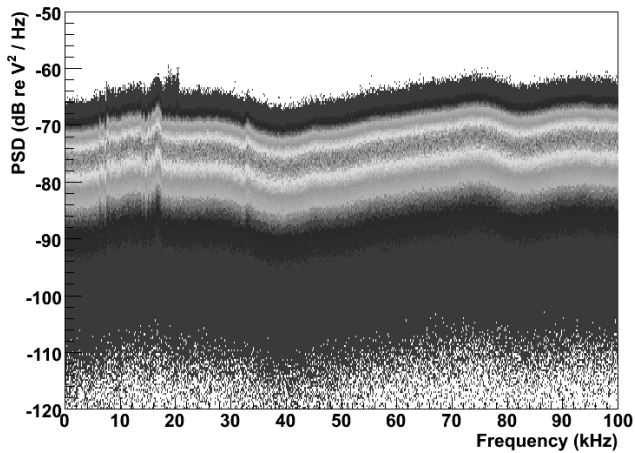}
  \caption{Long-term distribution of power spectral densities for
    string C, sensor 7, channel 2 (400~m).}
  \label{fig:psd}
\end{figure}

Figure \ref{fig:psd} shows the distribution of PSD values in each
frequency bin for sensor C7(2) (String C, sensor 7, channel 2) which
is at a depth of 400~m. The graph includes all available data from 1
January 2008 to 7 May 2008.  A comparison of all sensors shows that
the spectral shape of the noise is consistent for all three channels
within one sensor housing but differs strongly between different
sensors. This is an indication that the spectral shape is determined
mostly by the steel housing of the sensor, and the three channels in
one sensor are strongly coupled by the housing and a central preload
screw. This effect will be studied further by using the second
generation sensors that were installed on the fourth SPATS
string. These include SPATS type steel sensors with modified coupling
of the piezo element to the housing and HADES sensors
\cite{Semburg:2008} comprised of a piezo element and pre-amplifier
enclosed in resin.

\begin{figure}[ht]
  \centering
  \includegraphics[width=0.5\textwidth]{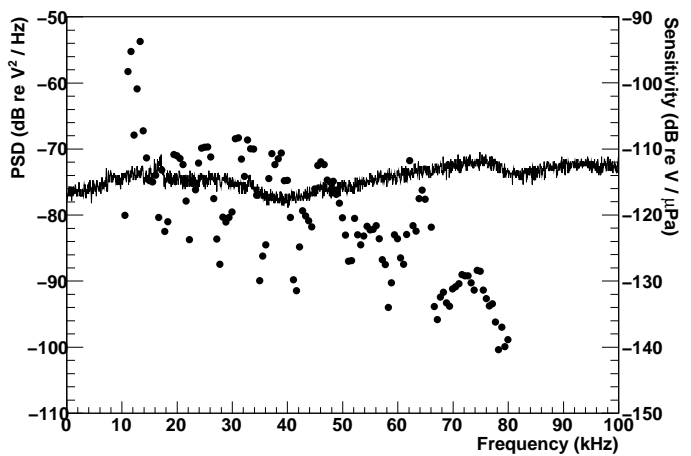}
  \caption{Mode (solid line) calculated per frequency bin of the
    distribution shown in Figure \ref{fig:psd}. Superimposed is the
    sensitivity of the sensor measured in water scaled by a factor of
    1.4 (dots, cf.~text).}
  \label{fig:psd_reduced}
\end{figure}

To reduce the data further the distribution of PSD values in each
frequency bin is evaluated. In Figure \ref{fig:psd_reduced} the mode
for each frequency bin is shown, which is a measure for the most
common noise level.

Also shown in Figure \ref{fig:psd_reduced} is the sensitivity measured
in water scaled by a factor of 1.4 as discussed in Section
\ref{sec:calibration}. Since the sensitivity decreases strongly at
high frequencies, only the frequency range from 10~kHz to 50~kHz is
considered for further analysis.

\begin{figure}[ht]
  \centering
  \includegraphics[width=0.5\textwidth]{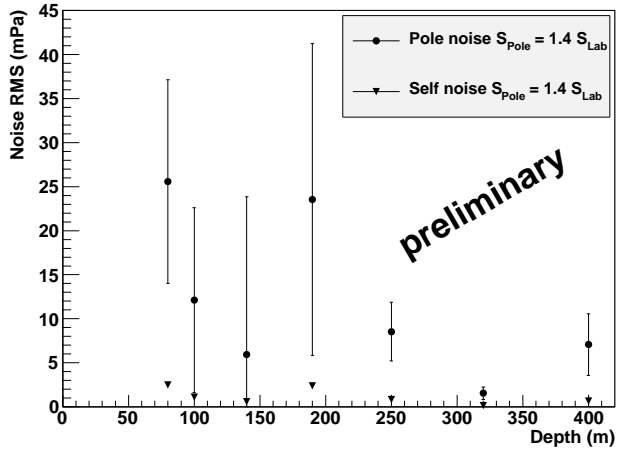}  
  \caption{Noise RMS integrated from 10~kHz to 50~kHz as function of
    sensor depth. The error bars only describe sensor to sensor
    variations.}
  \label{fig:rms_depth}
\end{figure}

After the scaled sensitivity is applied to the PSD mode, the resulting
pressure PSD is integrated over the relevant frequency range from
10~kHz to 50~kHz to determine the absolute noise level RMS. Figure
\ref{fig:rms_depth} shows the mean noise level as function of sensor
depth. The error bars represent the standard deviation of the values
measured in the up to 12 sensor channels available at each depth.

It can be seen that both the noise level and fluctuations are high in
the upper 200~m of the SPATS setup. In this region, the firn, a
transition takes place from a snow/air mixture to compact bulk
ice. Below the firn, the noise conditions become more stable and an
average noise level below 10~mPa in the frequency range relevant for
acoustic neutrino detection (10~kHz to 50~kHz) is derived.

\section{Conclusions}
\label{sec:conclusions}

The South Pole Acoustic Test Setup is successfully recording valuable
data to study the acoustic properties of deep South Pole ice with
respect to the feasibility of acoustic UHE neutrino detection for one
and a half years. We have shown that the level of background noise,
which is a critical parameter for the construction of an acoustic
neutrino telescope, is stable up to the time scale of one year. A
first estimation of the absolute noise level at depths larger than
200~m indicates values below 10~mPa integrated over the relevant
frequency range for neutrino detection from 10~kHz to 50~kHz. This
value is of the same order of magnitude as what is observed in marine
studies at good weather conditions \cite{Kurahashi:2007,
  Aiello:2008}. However, this result is subject to systematic
uncertainties which are under study. These include the change of
sensor sensitivity under the combined exposure to low temperatures and
high ambient pressure and the influence of the freshly frozen ``hole
ice'' the sensors are emerged in.

\section*{Acknowledgements}

We gratefully acknowledge the hospitality of the National Science
Foundation Amundsen-Scott South Pole Station.

This work was supported by the German Ministry for Education and
Research.

\end{document}